%% file: article.tex
\documentstyle[aps,12pt,epsf]{revtex} 

\def\PR #1 #2 #3 {Phys.~Rev.~{\bf #1}, #2 (#3)}
\def\PRL #1 #2 #3 {Phys.~Rev.~Lett.~{\bf #1}, #2 (#3)}
\def\PRD #1 #2 #3 {Phys.~Rev.~D~{\bf #1}, #2 (#3)}
\def\PLB #1 #2 #3 {Phys.~Lett.~{\bf B#1}, #2 (#3)}
\def\NPB #1 #2 #3 {Nucl.~Phys.~{\bf B#1}, #2 (#3)}
\def\RMP #1 #2 #3 {Rev.~Mod.~Phys.~{\bf #1}, #2 (#3)}
\def\ZPC #1 #2 #3 {Z.~Phys.~C~{\bf #1}, #2 (#3)}
\begin{document} 
\pagestyle{plain} 
\begin{titlepage} 
%\begin{flushright} 
%PKU-TH-98-62\\ 
%\end{flushright} 
\vspace{.5cm} 
 
\begin{center} 
{\Large Supersymmetric Higgs Bosons Discovery Potential at Hadron Colliders through $bg$ channel } 
 
\vspace{.2in} 
 Chao Shang  Huang$^a$ and Shou Hua Zhu$^{b,a}$  \\
 $^a$ Institute of Theoretical Physics, Academia Sinica, P. O. Box 
 2735, Beijing 100080, P. R. China \\
 $^b$ CCAST (World Laboratory), P. O. Box 8730, Beijing 100080, P. R. China \\ 
\end{center}

\begin{footnotesize} 
\begin{center}\begin{minipage}{5in} 
\baselineskip=0.25in 
\vspace{3cm}
\begin{center} ABSTRACT \end{center} 
We explore the discovery potential of the supersymmetric Higgs bosons through 
$bg$ channel at Tevatron and LHC. Compared with the process of $qq' 
\rightarrow WH$
, this channel is more advantigeous to
 finding the supersymmetric Higgs bosons at Tevatron if $\tan\beta$ is larger
than ten.
\end{minipage}\end{center} 
\end{footnotesize} 
\vfill 
 
 
\end{titlepage} 
 
\newpage 
 One of the most important physics goals for future high energy
physics is the discovery of the Higgs boson. Recent direct search 
in the LEP2 experiments of running at $\sqrt{s}=183$ Gev via the
$e^+e^-\rightarrow Z^*H$ yields a lower bound of $\sim 89.9$ Gev on 
the standard model (SM) Higgs mass \cite{a01}. Next year's running at $192$ Gev  
will explore up to a Higgs boson mass of about $96$ Gev 
\cite{a02}. 
After LEP2 the search for the Higgs particles will be continued
at the CERN Large Hadron Collider (LHC) for Higgs boson masses
up to the theoretical upper limit. 
Before the LHC comes into operation it is worth considering 
whether the Higgs boson can be discovered from the existing hadron 
collider, the Tevatron. Much study has been made in the detection 
of a Higgs boson at the Tevatron \cite{a03}. In Ref. \cite{a02},
it was pointed out that if the Higgs boson is discovered at LEP2, it should
be observed at the Tevatron's Run II with CM energy $\sqrt{s}=2$ Tev and
an integrated luminosity $\sim 10 fb^{-1}$, through the production 
subprocess $q\bar q'\rightarrow WH$, followed by $W \rightarrow \ell\bar 
\nu$ and $H\rightarrow b\bar b$, and if the Higgs boson lies beyond the reach
of LEP2, $m_H \geq (95-100)$ Gev, then a $5-\sigma$ discovery will be
possible in the above production sub-process in a future Run III with
an integrated luminosity $30 fb^{-1}$ for masses up to $m_H \approx
125$ Gev. However, we notice that this channel can't work for large $\tan\beta$ \cite{a07}. 
Recently,  Ref. \cite{Han} has studied the Higgs boson discovery potential of
the process $gg\rightarrow H$~\cite{ggmn} at Tevatron, and found that the SM-like
 Higgs boson could
be found if its mass lie in the range of $135$ to $180$ Gev. 
In literatures, there are also many works \cite{new} discussing 
Higgs bosons discovery abilities with b quarks at hardron colliders. 
For examples,
in the first reference of Ref. \cite{new}, the process
$P\bar P \rightarrow b \bar b H X$, in which the actual physical 
subprocess of the inclusive rate of Higgs production
associated with bottom quarks is $gg
\rightarrow b\bar b H$, has been examined. 
In this paper we examine the Higgs-bottom
association production $p\bar p\rightarrow b H (\bar b H) X$ in which
the actual physical subprocess is $bg\rightarrow b H$. It is evident
that this process is different from $p\bar p\rightarrow b\bar b H X$ in
tagging only one b quark in our case.\\

As we know, the distributions of the sea b-quark and gluon grow 
rapidly for small x region, when $x < 0.1 $, the gluon distribution function
is far larger than u quark  distribution function and the same thing occur
for sea b quark when $x < 0.01 $, 
so Tevatron and LHC are good places to examine the bg channel. \\

It is well-known that the couplings of CP-even neutral Higgs bosons to
down-type quarks in supersymmetric (SUSY) models are given by \cite{hh}
\begin{eqnarray}
\frac{-igm_D}{2m_W}\frac{cos\alpha}{cos\beta} ~~~~~~~~~~~~\mbox{ for }  H^0 D\bar {D}\\
\frac{-igm_D}{2m_w}\frac{sin\alpha}{cos\beta} ~~~~~~~~~~~~\mbox{for } h^0 D\bar{D}.
\end{eqnarray}
When tan$\beta \geq 35$ the  couplings of $H^0, h^0$ to b quark can be as 
large as those to t quark. Therefore, it is possible to discover SUSY Higgs
bosons, in particular for large tan$\beta$, at Tevatron through the bg
channel. \\

Including radiative corrections, the mixing angle $\alpha$ in eqs. (1,2) is
determined by
\begin{eqnarray}
tan2\alpha=\frac{sin2\beta~(m^2_A + m^2_Z)-2 R_{12}}{cos2\beta~(m^2_A
- m^2_Z) + R_{22}-R_{11}},~~~~~~~~~~-\frac{\pi}{2} <\alpha\leq 0,
\end{eqnarray}
 where $R_{ij}$ are the radiative corrections to the mass matrix of the neutral Higgs
bosons in the $\{H_1^0, H_2^0\}$ basis and have been given in references ~\cite{oyy,cqw}. An
analysis of the couplings of Higgs bosons to vector bosons, up-type and down-type
quarks in both large $tan\beta$ and large $m_A$ limits has been performed ~\cite{lw}
and some numerical results for $\tan\beta$ = 1.5 and 30 in vanishing mixing case have
been given in ref.~\cite{add}.
For our purpose, we shall concertrate on the general analysis of the couplings of
Higgs bosons to down-type quarks, based on the results given in ref.~\cite{cqw}. In order
to simplify discussions we assume $m_Q=m_U=m_D=m_S$ and consider the following three
representative cases. \\
(I)The case $A_t=A_b=\mu=0$\\
There is no mixing between stops as well as between sbottoms in this case. 
It should be noted that this case is of only an academical excise ($\mu=0$ is
ruled out by chargino and neutralino searches at LEP2).
The leading
corrections come from stop-loop and can be written as
\begin{eqnarray}
R_{11}=R_{12}=0,\\
R_{22}=\frac{3 G_F}{\sqrt{2}\pi^2}\frac{m_t^4}{\sin^2\beta} \log (1+ \frac{m_
S^2}{m_t^2} ),
\end{eqnarray}
where terms of order $\frac{m^2_Z}{(m^2_S+m^2_i)}$ (i=t, b) or $\frac{m^2_b}{m^2_t}$
have been neglected.\\
(II)The case $\mu\not\! = 0, A_t=A_b=0$\\
The radiative corrections depend on tan$\beta$ strongly. A large mixing between 
sbottoms happens while the mixing between stops is still small if tan$\beta$ is
large and $\mu$ is not too small~\footnote{In supergravity models due to the radiative electroweak symmetry
breaking mechanism one usually has $|\mu|\geq M_{1/2}$ at electroweak scale~\cite{dn} so that
the condition is satisfied.}. With $\mu >$100 Gev, tan$\beta \geq m_t/m_b$, and in the range of $m_S$ from 
500 Gev to 1 Tev, $R_{12}\sim R_{11}\sim$ a few thousandth of $R_{22}$.\\
(III)The case $\mu\sim A_t\sim A_b\not\! = 0$\\
There is a large mixing between stops. The mixing between sbottoms is
large if tan$\beta$ is large. In this case, for $\mu>$100 Gev and tan$\beta \geq m_t/m_b$, the 
radiative corrections to non-diagonal matrix element, $R_{12}$, can reach
more than ten percents of the raditive corrections to the diagonal matrix
element, $R_{22}$, while the radiative corrections to the another diagonal matrix
element, $R_{11}$, is still far smaller than $R_{22}$.

We calculate the cross sections of $bg \rightarrow bh^0$ and $bg \rightarrow bH^0$ in
all above three cases for different tan$\beta$. 
Through the paper, $m_A$ and $\tan\beta$ are choosen as input parameters.
The loop corrected masses of Higgs bosons $h^0$ and $H^0$ \cite{cqw} are
used in calculations.
The numerical results are given in 
Figs.1-2. 

In Fig. 1, we show the cross sections of the processes $bg \rightarrow bh^0$
and $bg \rightarrow bH^0$ for case (I) assuming $m_S=1$ Tev. 
From these curves, we can see that in a very wide region of $m_H$, the cross sections 
are much larger than that in SM, and can reach dozens of pb at Tevatron and $10^3$ pb
at LHC for large $\tan\beta$, which is due to  
 the enhancement of the couplings of $h_0-b-\bar b$ and $H-b-\bar b$
compared to the SM case. Compared with the $q\bar q' \rightarrow WH$ channel, 
the $bg\rightarrow bH$ channel is more advantageous to searching for SUSY Higgs bosons
if $\tan\beta$ is larger than 10, because for the $q\bar q' \rightarrow WH$ channel
the cross sections for the supersymmetric Higgs bosons are always smaller 
than the SM case, especially for large $\tan\beta$, which is
due to the suppression of $\sin(\beta-\alpha)$ \cite{a07}, and the
cross sections at most reach $1$ pb at Tevatron for the interesting
mass region of $95 - 125 $ Gev. Compared with the gluon-fusion mechanism $gg
\rightarrow H$ which is the dominant mechanism for neutral Higgs boson productions
at LHC for small and moderate values of tan$\beta$, the $bg\rightarrow bH$ channel can
compete with it at Tevatron and LHC if $\tan\beta\geq 35$. One can also see from the 
figure  that when the mass of the lightest Higgs boson 
approaches its upper limit,
the cross sections come back to the SM case, which is due to the reason
that the couplings of the lightest Higgs boson is the same as
the SM case when its mass approaches upper limit.  

From our numerical results, we find that the cross sections in the cases (II) and (III) 
are  similar to those in the case (I) in most of range of Higgs masses (below 120 Gev
for $h^0$ mass and above 140 Gev for $H^0$ mass), except in a narrow range around 130 
Gev where the cross sections in the case (III) are significantly 
different from those in the case (I). As an examples, in Fig. 2 we show the cross 
sections of three cases in the narrow mass region at Tevatron. It is evident from the
figure that the upper limit of 
$h^0$ mass for case (III) increases by about two  Gev compared to case (I), while much 
less variations appear for case (II). 

We did similar calculations for $m_S$ = 0.5 Tev. And results are that the cross 
sections have little changes while the shift of the upper limlit of $h^0$ mass is
significant.

Fig. 3 and 4 are devoted to the processes $bg \rightarrow b A^0$ and 
$bg \rightarrow t H^-$. Since the coupling of psedoscalar Higgs boson to 
b quark is proportional $\tan\beta$, the cross sections
increase quadraticly with the increment of $\tan\beta$ and can reach several dozens
 pb at Tevatron.
We notice that the cross section of the charged Higgs boson for $\tan\beta=10$ is smaller
than those of $\tan\beta=2$ and $40$, which is the consequence
of the competitive between the couplings $m_t/\tan\beta$ and $m_b \tan\beta$. 
  
To summarize, as a complementary process of $qq' \rightarrow WH$ 
and $gg \rightarrow H$, $bg$ channel could be very important in finding 
the supersymmetric Higgs bosons at Tevatron and LHC. In particular, it is possible to
find the SUSY neutral Higgs bosons at Tevatron via bg channel if $\tan\beta\geq 10$.
Anyway, the real Monte Carol simulation including QCD and electro-weak 
corrections 
is needed, and will give the further information for experiments.

\section*{Acknowledgments} 
This research was supported in part by the National Nature Science
Foundation of China and the post doctoral foundation
of China. S.H. Zhu gratefully acknowledges the
support of  K.C. Wong Education Foundation, Hong Kong.

%\newpage 

\newpage
\input{fig}

\end{document}

%% file: fig.tex
%\documentstyle[aps,12pt,epsf]{revtex}
%\begin{document}

\begin{figure}
\epsfxsize=15 cm
\centerline{\epsffile{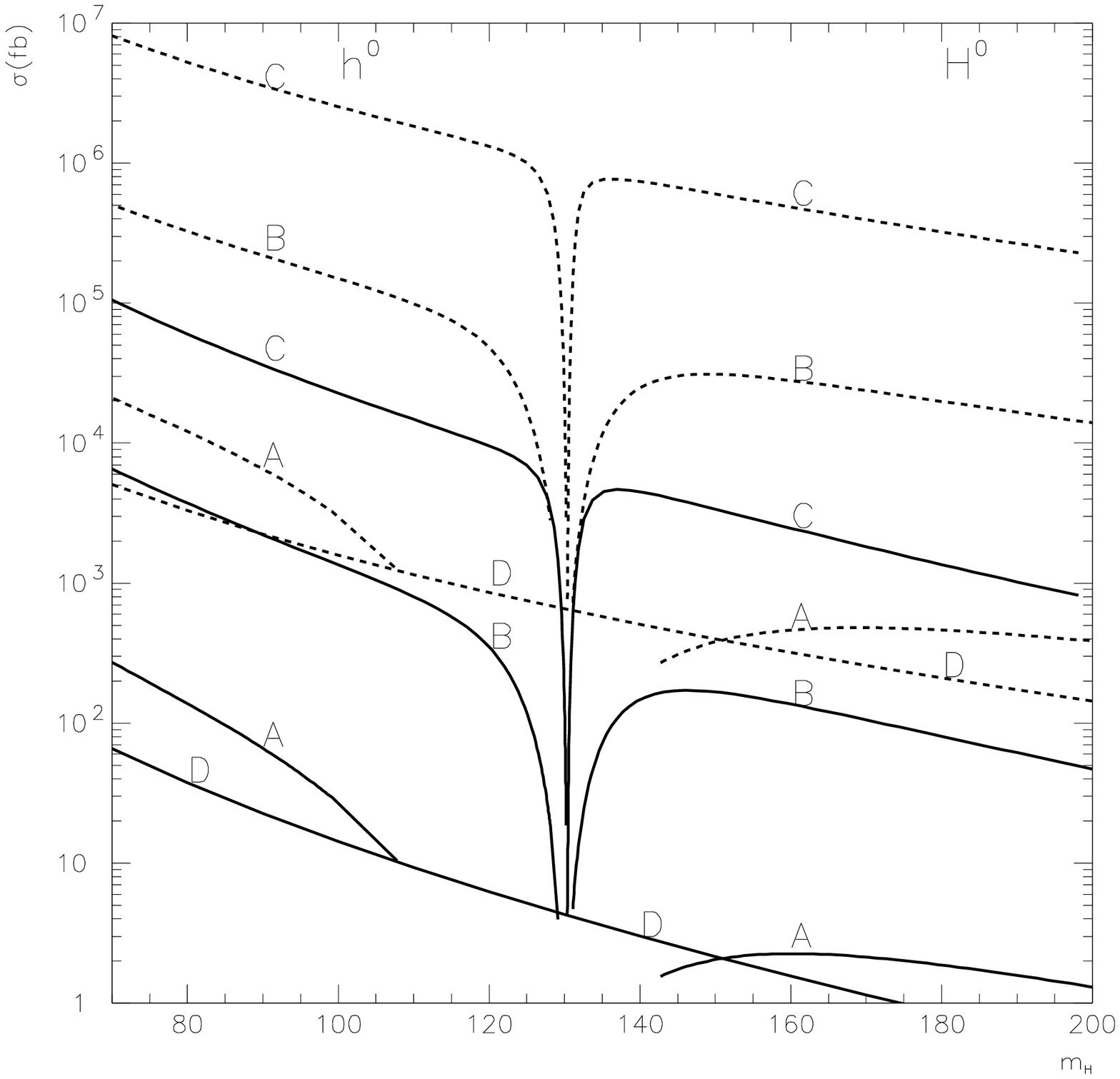}}
\caption[]{
The total cross sections versus $m_{H}$ for case (I), where
$m_S=1$ Tev, and
the solid and dashed lines represent the results at Tevatron and LHC,
and A, B, C  and D represent $\tan\beta=2, 10, 40$
and in the SM, respectively.
}
\end{figure}

\begin{figure}
\epsfxsize=15 cm
\centerline{\epsffile{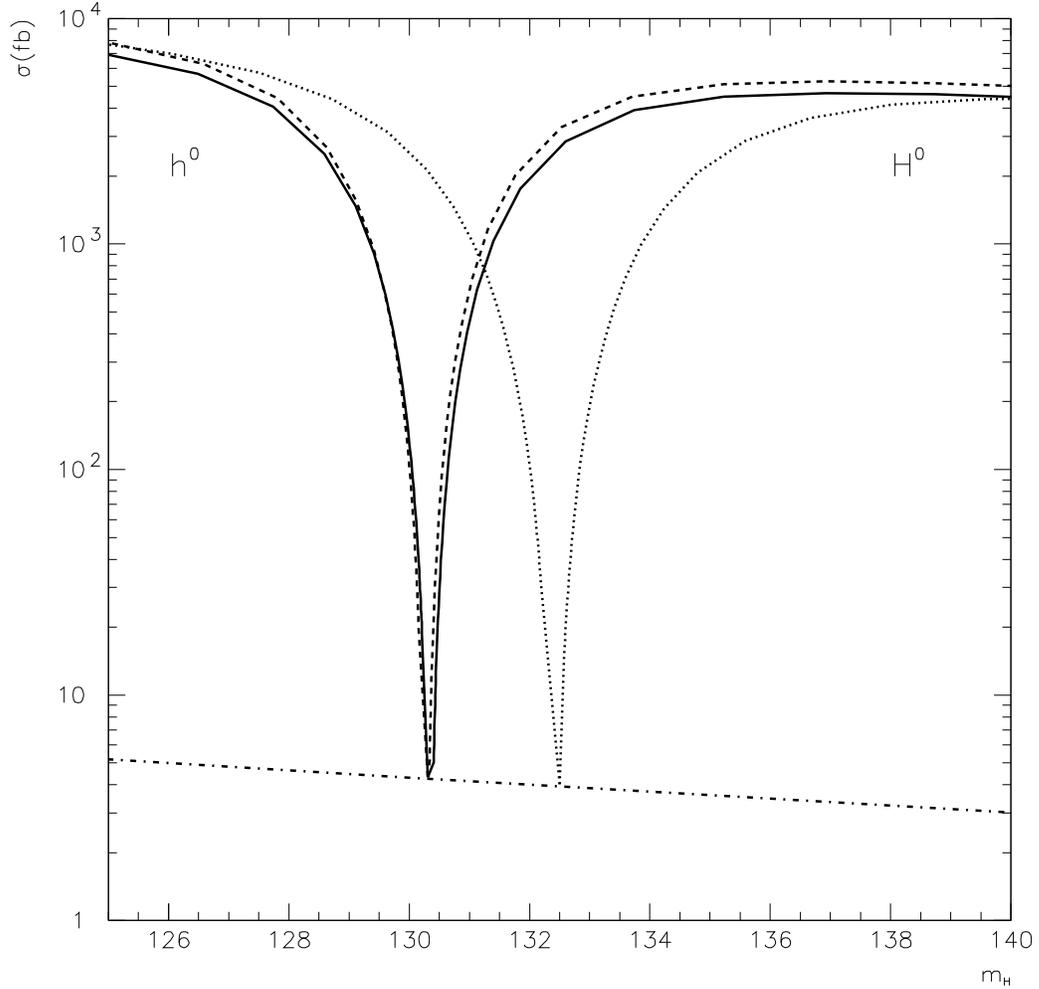}}
\caption[]{
The total cross sections versus $m_{H}$, where
$m_S=1$ Tev, 
$\tan\beta=40$, and
the solid, dashed, dotted and dot-dashed lines represent 
the results for case (I), (II), (III) and in SM, respectively.
For case (II), $A_t=A_b=0$ and $\mu=-500$ Gev; for case (III),
$A_t=A_b=\mu=-500$ Gev.
}
\end{figure}

\begin{figure}
\epsfxsize=15 cm
\centerline{\epsffile{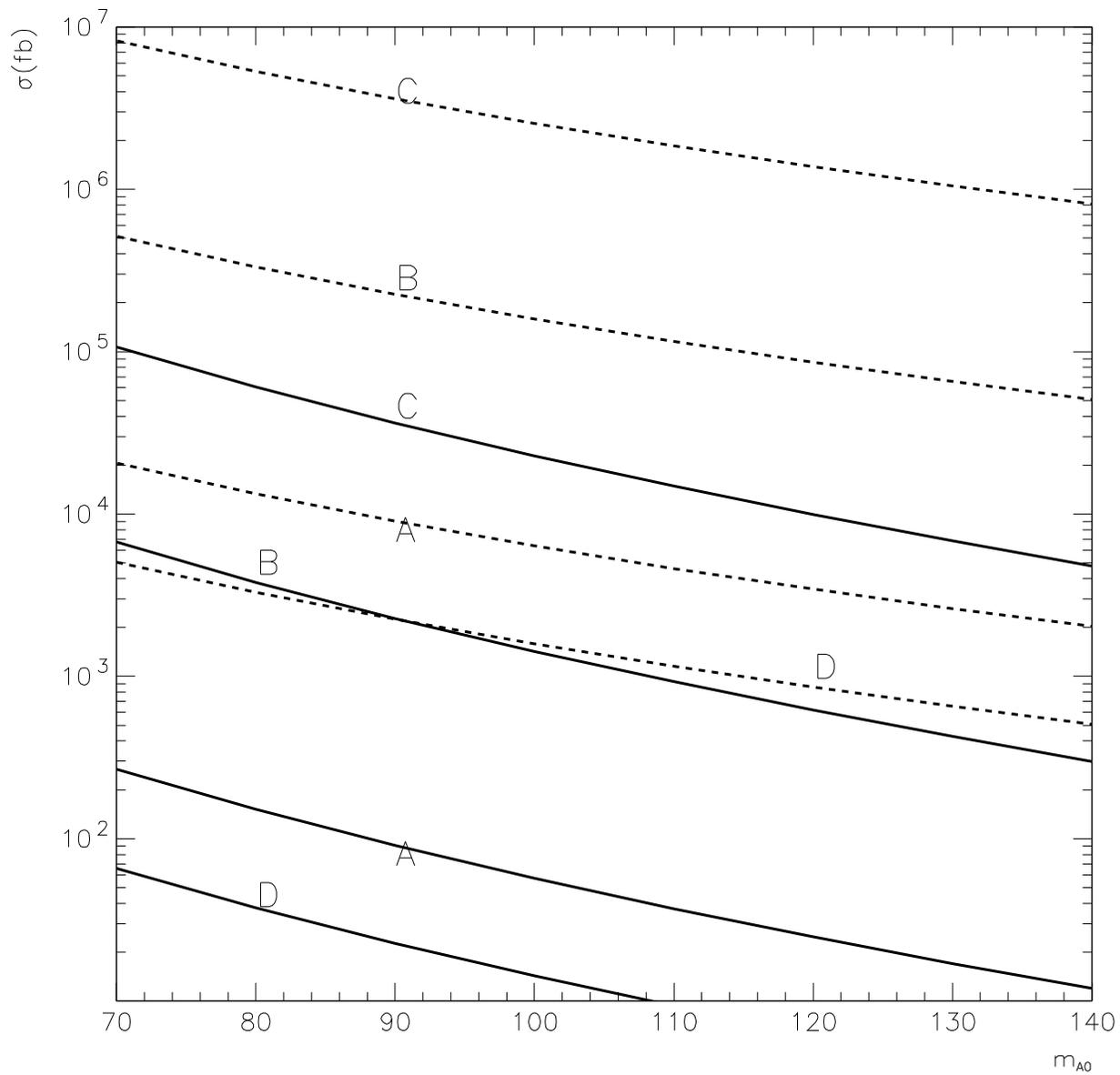}}
\caption[]{
The total cross sections versus $m_{A0}$, where
the solid and dashed lines represent the results at Tevatron and LHC,
respectively, and A, B,  C and D represent $\tan\beta=2, 10, 40$ and
in the SM.
}
\end{figure}

\begin{figure}
\epsfxsize=15 cm
\centerline{\epsffile{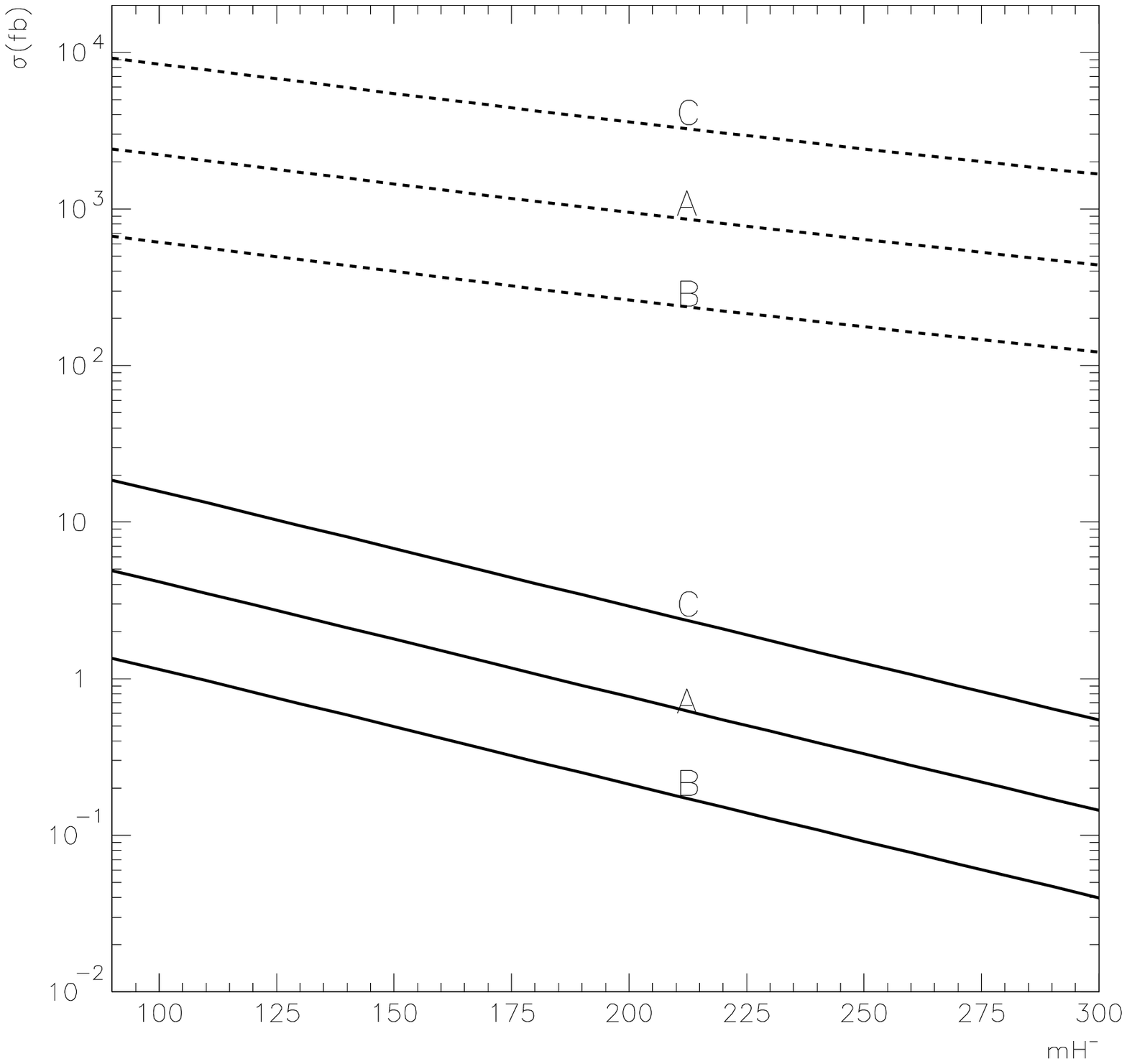}}
\caption[]{
The total cross sections versus $m_{H^-}$, where
solid and dashed lines represent the results at Tevatron and LHC,
respectively, and A, B and C represent $\tan\beta=2, 10, 40$,
respectively.}
\end{figure}

%\end{document}

%% file: article.bbl
\begin{thebibliography}{Espinosa} 

\bibitem{a01} 
P. McNamara, {\it ICHEP '98}, Vancouver, July 1998.

\bibitem{a02} 
C. Quigg, FERMILAB-CONF-98/059-T, 
hep-ph/9802320.
 
\bibitem{a03} A. Stange, W. Marciano, and S. Willenbrock, 
{Phys. Rev.} {\bf D49}, 1354 (1994); {\sl ibid.} {\bf D50},4491 (1994); 
S. Kim, S. Kuhlman, and W.Yao, CDF-ANAL-EXOTIC-PUBLIC-3904,  Oct. 96;
W.Yao, FERMILAB-CONF-96-383-E,  Jun. 96;
J. Womersley, D0 Note 3227, Apr. 97;
S. Parke, FERMILAB-CONF-97/335-T. 

\bibitem{a05} 
T. Han and S. Willenbrock, Phys. Lett. {\bf B273}, 167 (1990);\\ 
J. Ohnemus and W.J. Stirling, Phys. Rev. {\bf D47}, 2722 (1993);\\ 
H. Baer, B. Bailay, and J.F. Owens, Phys. Rev. {\bf D47}, 2730 (1993);\\ 
S. Smith and S. Willenbrock, Phys. Rev. {\bf D54}, 6696 (1996). 

\bibitem{a07} C.S. Li and S.H Zhu, Phys. Lett. {\bf B444},
224 (1998); Q.H. Cao, C.S. Li and S.H. Zhu,
 hep-ph/9810458.

\bibitem{Han} T. Han, R.J. Zhang, hep-ph/9807424. 

\bibitem{ggmn}H. Georgi, S. Glashow, M. Machacek and D. V. Nanopoulos, Phys. Lett.
{\bf 40}, 692 (1978).

\bibitem{new} 
D. Dicus, T. Stelzer, Z. Sullivan and S. Willenbrock, hep-ph/9811492;
Z.~Kunszt and F.~Zwirner, \NPB 385 3 1992;
 M.~Drees, M.~Guchait, and P.~Roy, \PRL 80 2047 1998 ;
Erratum {\bf 81}, 2394 (1998);
 M.~Carena, S.~Mrenna, and C.~Wagner, hep-ph/9808312;
 J.~Dai, J.~Gunion, and R.~Vega, \PLB 345 29 1995 ;
\PLB 387 801 1996;
 J.~Diaz-Cruz, H.-J.~He, T.~Tait, and C.-P.~Yuan,
\PRL 80 4641 1998;
C.~Balazs, J.~Diaz-Cruz, H.-J.~He, T.~Tait, and C.-P.~Yuan, hep-ph/9807349;
D.~Choudhury, A.~Datta, and S.~Raychaudhuri, hep-ph/9809552;
 C.~Kao and N.~Stepanov, \PRD 52 5025 1995;
V.~Barger and C.~Kao, \PLB 424 69 1998;


\bibitem{hh} J. Gunion, H. Haber, G. Kane and S. Dawson, The Higgs Hunter's Guide
(Addison-Wesley, Reading, MA, 1990). 

\bibitem{oyy} Y. Okada, M. Yamaguchi and T. Yanagida, Prog. Theor. Phys. {\bf 85},
1(1991); H. Haber and R. Hempfling, Phys. Rev. Lett. {\bf 66}, 1815(1991); J. Ellis,
G. Ridolfi and F. Zwirner, Phys. Lett. {\bf B257}, 83(1991).

\bibitem{cqw} M Carena, M. Quiros and C. Wagner, Nucl. Phys. {\bf B461}, 407(1996).

\bibitem{lw} W. Loinaz and J. Wells, hep-ph/9808287.
 
\bibitem{a06} 
H.E. Haber and G.L. Kane, {Phys. Rep.} {\bf 117}, 75(1985);\\ 
J.F. Gunion and H.E. Haber, Nucl. Phys. {\bf B 272}, 1 (1986). 
 
%\bibitem{a11} J. Gunion and A. Turski, Phys. Rev. {\bf D39 }, 2701 (1989) and {\bf D40} 
 %2333 (1990);\\ 
%J.R. Espinosa and M. Quiros, Phys. Lett. {\bf B266}, 389 (1991);\\ 
%M. Carena, M. Quiros, and C.E.M. Wagner, Nucl. Phys. {\bf B461}, 407 (1996). 
 
 
\bibitem{GRV} 
M. Gluck, E. Reya and A. Vogt, Z. Phys. C53, 127 (1992).

\bibitem{add} M. Spiral, CERN-TH/97-68 (hep-ph/9705337).

\bibitem{dn} M. Drees and M. M. Nojiri, Nucl. Phys. {\bf B369}, 54 (1992).

\end{thebibliography}
